\documentclass[12pt]{amsart}
\usepackage{amsmath}
\usepackage{tabularx}
\usepackage{tikz}
\usepackage{graphicx}  
\usepackage{adjustbox} 
\usetikzlibrary{positioning, calc}

\usepackage[hmargin=1.2in,vmargin=1.2in]{geometry}

\usepackage[linesnumbered,ruled,vlined]{algorithm2e}

\usepackage{amsmath,amsthm,amssymb,amsfonts,verbatim,}
\usepackage[colorlinks,
            linkcolor=blue,
            anchorcolor=blue,
            citecolor=blue
            ]{hyperref}
\usepackage{array}
\usepackage{multirow}

\usepackage{color}
\usepackage{booktabs}

\title{On Self-orthogonal MDS Codes}

\author{Shu Liu}
\address{National Key Laboratory of Wireless Communications, University of Electronic Science and Technology of China, Chengdu, China} \email{shuliu@uestc.edu.cn}

\date{}

\newtheorem{lemma}{Lemma}[section]
\newtheorem{theorem}[lemma]{Theorem}

\newtheorem{prop}[lemma]{Proposition}

\newtheorem{rem}[lemma]{Remark}

\newtheorem{conjecture}{Conjecture}

\newtheorem{main}{Main Theorem}
\newtheorem{open}{Open Problem}
\theoremstyle{remark}

\renewcommand{\epsilon}{\varepsilon}
\renewcommand{\le}{\leqslant}
\renewcommand{\ge}{\geqslant}

\newcommand{\vnote}[1]{}

\def\ZZ{\mathbb{Z}}
\def\PP{\mathbb{P}}

\def\F{\mathbb{F}}
\def \mC {\mathcal{C}}
\def \mA {\mathcal{A}}

\def \mA {\mathcal{A}}

\def \mC {\mathcal{C}}

\def \mL {\mathcal{L}}

\def \mX {\mathcal{X}}

\def\Pin{{P_{\infty}}}

\def \Xi {{X^{[i]}}}

\newcommand{\Ga}{\alpha}
\newcommand{\Gb}{\beta}
\newcommand{\Gg}{\gamma}

\newcommand{\Gl}{\lambda}    
\newcommand{\Go}{\omega}     \newcommand{\GO}{\Omega}

\def \ba {{\bf a}}
\def \bb {{\bf b}}
\def \bc {{\bf c}}

\def \bu {{\bf u}}
\def \bv {{\bf v}}
\def \bo {{\bf 0}}
\def\Supp {{\rm Supp }}
\def\supp {{\rm supp }}

\def\wt{{\rm wt}}

\def\RS{{\textsf{RS} }}

\def\res{{\rm res}}

\begin{document}
\maketitle

\begin{abstract} Both self-orthogonal codes and MDS codes are of great interest for  coding and combinatorial theory. In this paper we consider the existence problem of self-orthogonal MDS codes, i.e, codes satisfying self-orthogonality and achieving the Singleton bound.  More precisely, we investigate for which length $n$ and dimension $k$, there exist  $q$-ary $[n,k]$ self-orthogonal MDS codes. We completely solve this problem for even prime power.  In
odd characteristic, we establish existence results for $n=q+1, n|q$, and $n | (q-1)$,
together with constructions at additional lengths.

Our technique is to make use of the language of algebraic geometry codes. An advantage of using  the language of algebraic geometry codes is that the Euclidean dual code of a functional algebraic geometry code is a differential code which can be turn into a  functional  code again. Thus, by choosing an appropriate differential, one can give an explicit construction  of  self-orthogonal MDS codes.
\end{abstract}

\section{Introduction}
In literature, self-dual MDS codes have been extensively studied in recent years \cite{FF19,FLLL19,FLL21,GK02,GG08,G12,GKL08,HK05,HK06,JX17,KZT12,KL04a,KL04b,LLL19,TW16,Y19,YC15,ZF19}, while  self-orthogonal MDS codes have been less studied. Though a self-dual code of length $n$ can induce self-orthogonal codes for any dimension $k$ with $k\le n/2$. However, a self-dual MDS code of length $n$ does  not necessarily produce  an $[n,k] $ self-orthogonal MDS code for dimenion $k<n/2$. One counter-example given in {Subsection 4.1} is that for an odd prime $q$, one can explicitly construct a $q$-ary $[q+1,(q+1)/2]$ self-dual MDS code $\mC$, while we also show non-existence of   subcodes of $\mC$ that are $q$-ary {$[q+1,(q-1)/2]$} self-orthogonal MDS codes by assuming the Main MDS conjecture (see Proposition \ref{prop:4.10}).

The main purpose of this paper is to study the existence problem of self-orthogonal MDS codes.  
\subsection{Known results} 
In the literature, there are tremendous amount of papers studying either self-orthogonal codes or  MDS codes. However, a few papers have studied self-orthogonal MDS codes. In \cite{GK02,HK05,HK06}, the authors make use of combinatorial methods, i.e., orthogonal designs to construct  self-dual MDS codes. Some other constructions of self-dual MDS codes are given in \cite{KL04a,KL04b} via building up techniques; and in \cite{KZT12,TW16,YC15} via constacyclic codes; and in \cite{FF19,FLL21,JX17,LLL19,TW16,ZF19} via generalized Reed-Solomon (GRS) codes.  

A class of self-orthogonal MDS codes is given in \cite{FLL21}  via GRS codes. Precisely speaking, it was shown in \cite{FLL21} that if $q=r^2$, $1\le s\le \frac{r+1}2$, $1\le t\le \frac{r-1}2$ and $n=s(r-1)+t(r+1)$, then there exist $q$-ary $[n,k]$  self-orthogonal MDS codes for any $1\le k\le \frac n2$. Thus, the paper \cite{FLL21} focus only on finite fields $\F_q$ with $q$ being a perfect square.  While our paper considers arbitrary finite fields by distinguishing even and odd characteristics. 

\subsection{Our contributions and techniques}
In this paper, we provide explicit constructions of self-orthogonal MDS codes for various parameters. In particular, by assuming the Main MDS conjecture (see Subsection 2.1),  we completely solve the existence problem of self-orthogonal MDS codes for even characteristic, namely we have the following result.
\begin{main}
Let $q$ be a power of $2$, then we have
\begin{itemize}
    \item[{\rm (i)}] For every $n\le q+1$ and $k\le n/2$ except for $[n,k]=[q+1,q/2]$, there exists a $q$-ary $[n,k]$ self-orthogonal MDS codes.
     \item[{\rm (ii)}] There do not exist any $q$-ary $[q+1,q/2]$ self-orthogonal MDS codes if the Main MDS conjecture holds.
   {
   \item[{\rm(iii)}] A linear self-orthogonal \([q+2,3]_q\) MDS code exists if and only if \(q\ge16\).}
\end{itemize}
\end{main}

For an odd prime power $q$, we also completely solve the existence problem of self-orthogonal MDS codes for the cases where the code length $n$ satisfies $n=q+1$, $n|q$ and $n|(q-1)$ except for existence of $q$-ary $[q-1,(q-1)/2]$ self-dual codes with $q\equiv1\pmod{4}$, namely we have the following results.

\begin{main}\label{main:2}
Let $q$ be an odd prime power, and let $1\le k\le\lfloor n/2\rfloor$. 
\begin{itemize}
\item[{\rm(i)}] Length $n=q+1$. An \([n,k]_q\) self-orthogonal MDS code exists for  \(k\ne(q-1)/2\).   For $k=(q-1)/2$, no $[n,k]_q$  self-orthogonal MDS codes exist for prime $q$ or $q=9$. Furthermore, for all prime power $q$, no extended generalized Reed–Solomon codes with parameters $[n,(q-1)/2]_q$ are self-orthogonal.
\item[{\rm(ii)}] Lengths \(n\mid q\). An \([n,k]_q\) self-orthogonal MDS code exists for every \(1\le k\le\lfloor n/2\rfloor\).
\item[{\rm(iii)}] Lengths \(n\mid(q-1)\). An \([n,k]_q\) self-orthogonal MDS code exists for every \(1\le k<n/2\). When \(n\) is even and \(k=n/2\), the following hold.
\begin{itemize}
    \item If \(q\equiv3\pmod4\), no such a self-orthogonal MDS code exists.
   \item  If \(q\equiv1\pmod4\) and \(n<q-1\), there is an $[n,n/2]_q$ self-dual MDS code.
   \end{itemize}
       \end{itemize}
\end{main}

\begin{conjecture} Let \(q=p^m\ge25\), where \(p\) is an odd prime and \(m\ge2\). Then no linear Euclidean self-orthogonal \([q+1,(q-1)/2]_q\) MDS code exists.
\end{conjecture}

\begin{rem}{\rm 
By Main Theorem \ref{main:2} the only remaining unsolved case for $n|(q-1)$ is whether there are $q$-ary self-dual MDS codes for $q\equiv 1\pmod{4}$ and $n=q-1$. This case has no definite answer: for instance, there exist no 
$q$-ary self-dual MDS codes of length  $q-1$ for $q=5,13$, etc., while such codes do exist  
for $q=9$ (see \cite{GG08}). 
}\end{rem}

The above remark leaves the following open problem.
\begin{open}
For which odd prime power $q$ with $q\equiv1\pmod{4}$,  do there exist $[q-1,(q-1)/2]_q$   self-dual MDS codes?  
\end{open}

In the case of odd prime power, we also show existence of self-orthogonal MDS codes for some other parameter regimes in Subsection 4.2. 

In summary, we present the following table for code parameters for which there exist and nonexist self-orthogonal MDS codes.

\begin{table}[htbp]
\centering
\caption{Existence of linear self-orthogonal MDS codes.}
\label{tab:self-orthogonal-mds}
\small
\renewcommand{\arraystretch}{1.18}
\setlength{\tabcolsep}{5pt}
\begin{tabularx}{\textwidth}{@{}
  >{\raggedright\arraybackslash}p{0.20\textwidth}
  >{\raggedright\arraybackslash}X
  >{\raggedright\arraybackslash}p{0.21\textwidth}
  >{\raggedright\arraybackslash}p{0.21\textwidth}
@{}}
\toprule
Field size \(q\) & Length \(n\) and conditions
  & Dimension \(k\) & Source \\
\midrule
Even
  & \(2\le n\le q\)
  & \(1\le k\le\lfloor n/2\rfloor\)
  & \textbf{Theorem~4.1}; [6] at \(k=n/2\) \\

Even
  & \(n=q+1\)
  & \(1\le k\le q/2-1\)
  & \textbf{Theorem~4.3} \\

\(q=2^e,\ e\ge4\)
  & \(n=q+2\)
  & \(k=3\)
  & \textbf{Theorem~4.7} \\
\midrule
Odd
  & \(n=q+1\)
  & \(1\le k\le(q-3)/2\), or \(k=(q+1)/2\)
  & \textbf{Theorem~4.8}; [6, 11] at \(k=(q+1)/2\) \\

Odd
  & \(n\mid q\)
  & \(1\le k\le\lfloor n/2\rfloor\)
  & \textbf{Theorem~4.12} \\

Odd
  & \(n\mid(q-1)\)
  & \(1\le k\le\lfloor(n-1)/2\rfloor\)
  & \textbf{Theorem~4.13} \\

\(q\equiv1\pmod4\)
  & \(n\) even, \(n\mid(q-1)\), and \(n<q-1\)
  & \(k=n/2\)
  & Yan [22] \\

\(q=9\)
  & \(n=8\)
  & \(k=4\)
  & Grassl--Gulliver [6] \\

Odd
  & \((n+m)\mid q\) for an integer \(m\ge1\)
  & \(1\le k\le\lfloor(n-m)/2\rfloor\)
  & \textbf{Theorem~4.14(i)} \\

Odd
  & \((n+m)\mid(q-1)\) for an integer \(m\ge1\)
  & \(1\le k\le\lfloor(n-m-1)/2\rfloor\)
  & \textbf{Theorem~4.14(ii)} \\

\(q=p^e\equiv1\pmod4\), \(p\) prime
  & \(n=p^\ell+1\), \(0\le\ell<e\)
  & \(1\le k\le n/2\)
  & \textbf{Theorem~4.16}; [2] at \(k=n/2\) \\
\midrule
\(q=r^2\), \(r\) an odd prime power
  & \(n=s(r-1)+t(r+1)\),\newline
    \(1\le s\le(r+1)/2\), \(1\le t\le(r-1)/2\)
  & \(1\le k\le n/2-1\)
  & Fang--Liu--Luo [4, Theorem~2] \\

\(q=r^e\), \(r\) an odd prime power, \(e\) even
  & \(n=2tr^\ell\),\newline
    \(1\le t\le(r-1)/2\), \(0\le\ell<e\)
  & \(1\le k\le n/2\)
  & Fang--Fu [2] \\

\(q=r^e\), \(r\) an odd prime power, \(e\) even
  & \(n=(2t+1)r^\ell+1\),\newline
    \(0\le t\le(r-1)/2\), \(0\le\ell<e\)
  & \(1\le k\le n/2\)
  & Fang--Fu [2] \\

\(q\equiv1\pmod4\)
  & \(n\) even and \(4^n n^2\le q\)
  & \(1\le k\le n/2\)
  & Jin--Xing [11] \\
\bottomrule
\end{tabularx}

\par\smallskip
\begin{minipage}{\textwidth}
\footnotesize
All length and dimension parameters are integers, with \(n\ge2\) and \(k\ge1\).
The endpoint \(k=n/2\) gives a Euclidean self-dual MDS code.
In the additional families from [2] and [11], the smaller dimensions
follow by taking nested GRS subcodes of the self-dual construction.
\end{minipage}
\end{table}

\begin{table}[htbp]
\centering
\caption{Nonexistence results for linear Euclidean self-orthogonal MDS codes.}
\label{tab:nonexistence-self-orthogonal-mds}
\small
\renewcommand{\arraystretch}{1.18}
\setlength{\tabcolsep}{5pt}
\begin{tabularx}{\textwidth}{@{}
  >{\raggedright\arraybackslash}p{0.18\textwidth}
  >{\raggedright\arraybackslash}X
  >{\raggedright\arraybackslash}p{0.32\textwidth}
@{}}
\toprule
Field size \(q\) & Parameters \((n,k)\) and scope & Source \\
\midrule
\(q\equiv3\pmod4\)
  & \(n\equiv2\pmod4,\quad k=n/2\)
  & Pless [18]; \textbf{Proposition~4.17} \\

Prime \(q\)
  & \(n>q+1,\quad 2\le k\le n-2\)
  & Ball [1] \\

\(q=5\)
  & \(n=4,\quad k=2\)
  & Remark~1.1 \\

\(q=2,4\)
  & \(n=q+1,\quad k=q/2\)
  & Proposition~4.4 \\

Even \(q\ge8\)
  & \(n=q+1,\quad k=q/2\)
  & \textbf{Theorem~4.5}\\

\(q=4,8\)
  & \(n=q+2,\quad k=3\)
  & Proposition~4.6; \textbf{Theorem~4.7} \\

Even \(q\ge16\)
  & \(n=q+2,\quad k=3\), regular hyperovals
  & \textbf{Theorem~4.7} \\

Odd \(q\)
  & \(n=q+1,\quad k=(q-1)/2\), extended GRS codes
  & \textbf{Theorem~4.9} \\

Odd prime \(q\), or \(q=9\)
  & \(n=q+1,\quad k=(q-1)/2\)
  & \textbf{Theorem~4.9} \\

Odd \(q\)
  & \(n=q+1,\quad k=(q-1)/2\), subcodes of
    self-dual \([q+1,(q+1)/2]_q\) MDS codes
  & \textbf{Proposition~4.10} \\
\bottomrule
\end{tabularx}
\end{table}
We make use of the language of algebraic geometry codes to present a framework for constructions of self-orthogonal MDS codes. Though the algebraic geometry codes from rational function fields are actually GRS codes (see Subsection 2.3), it is still more convenient to use the language of algebraic geometry codes due to the fact that the dual code of an algebraic geometry code can be easily defined via the same divisors. Furthermore, when we include ``the point at infinity" in the construction of GRS codes, it is more natural to characterize the dual codes via the language of algebraic geometry codes. Thus, to obtain a self-orthogonal GRS code via the language of algebraic geometry codes, one just need to construct certain differential (see Section 3 for the detail). 

More precisely, our framework works as follows. We choose a divisor $G$ and $n$ distinct rational places $P_1,\dots,P_n$ such that $P_i\not\in\Supp(G)$. Put $D=\sum_{i=1}^nP_i$ and consider the function code $\mC_L(D,G,\bu)$ for a vector $\bu=(u_1,\dots,u_n)\in(\F_q^*)^n$. Then the dual code of $\mC_L(D,G,\bu)$ is the differential code $\mC_\GO(D,G,\bu^{-1})$, where $\bu^{-1}=(u_1^{-1},\dots,u_n^{-1})$. Furthermore,  if $\eta$ is a differential satisfying $\nu_{P_i}(\eta)=-1$ for all $1\le i\le n$, then the dual code $\mC_\GO(D,G,\bu^{-1})$ of $\mC_L(D,G,\bu)$ is $\mC_L(D, D-G+(\eta), \bu^{-1}*{\bf res}(\eta))$, where ${\bf res}(\eta)=(\res_{P_1}(\eta), \dots,\res_{P_n}(\eta))$. Thus, by choosing a suitable differential $\eta$, one obtains  $\mC_L(D,G,\bu)\subseteq \mC_L(D, D-G+(\eta), \bu^{-1}*{\bf res}(\eta))$. Under this framework, we are able to construct various self-orthogonal MDS codes.

\section{Preliminaries}
In this section, we introduce self-orthogonal codes and MDS codes, algebraic geometry codes, rational function fields and generalized Reed-Solomon (GRS) codes.

\subsection{Self-orthogonal codes and MDS codes}
Let $q$ be a prime power and let $\F_q$ denote the finite field with $q$ elements. A $q$-ary $[n,k,d]$-linear code $\mC$ is an $\F_q$-subspace of $\F_q^n$ of dimension $k$ with minimum distance $d$. The well-known Singleton bound says that $d+k\le n+1$. A $q$-ary $[n,k,d]$-linear code is called  Maximum Distance Separable (MDS for short) if $k+d=n+1$. 

The {\bf Main MDS Conjecture} states: {\it for a nontrivial $q$-ary linear $[n, k]$ MDS code with $1<k<n-1$,, the code length $n$ is at most $q+1$, unless $q$ is even and $k=3$ or $k=q-1$, in which case $n\le q+2$.}

The dual code $\mC^\perp$ of $\mC$ is  given by $\{\bu\in\F_q^n:\; \langle \bu,\bc\rangle=0\; \mbox{for any}\ \bc\in\mC\}$, where $\langle \cdot,\cdot\rangle$ stands for the Euclidean inner product. $\mC$ is called self-orthogonal and dual, respectively,  if $\mC\subseteq\mC^\perp$ and $\mC=\mC^\perp$, respectively.

A $q$-ary $[n,k,d]$-linear code $\mC$ is called self-orthogonal MDS (self-dual MDS, respectively) if $\mC$ is both self-orthogonal and MDS (both self-orthogonal and MDS, respectively).

Let \(M\) be a full-row-rank \(k\times n\) generator matrix of a linear code . Then $\mC$ is Euclidean self-orthogonal if and only if \(MM^{\mathsf T}=0\). In particular, \(\dim C^\perp=n-k\), so a self-orthogonal code has \(k\le n/2\), with equality if and only if it is self-dual.

An \([n,k]\) code is MDS if and only if every \(k\) columns of a generator matrix are linearly independent. The dual of an \([n,k]\) MDS code is an \([n,n-k]\) MDS code. For \(1\le r<k\), shortening an MDS code on any \(r\) coordinates gives an \([n-r,k-r,n-k+1]\) MDS code. Shortening means taking the subcode of words that vanish on those coordinates and then deleting those coordinates. This operation preserves Euclidean self-orthogonality.

A change of row basis preserves the condition \(MM^{\mathsf T}=0\), since \((AM)(AM)^{\mathsf T}=A(MM^{\mathsf T})A^{\mathsf T}\). Coordinate permutations also preserve Euclidean self-orthogonality. For nonzero column scalars \(\lambda_i\), the Gram matrix after scaling is \(M\operatorname{diag}(\lambda_1^2,\ldots,\lambda_n^2)M^{\mathsf T}\).

There are some propagation rules in coding theory. Some codes can be constructed with  smaller sizes  or shorter lengths from codes of larger sizes or longer lengths. This work details one such rule employed for our code construction.
\begin{lemma}(see \cite{LX2004})~\label{lem.P1}
Suppose there is an $q$-ary $[n,k,d]$-linear code over $\F_q$. Then, there exits an $[n-r,k-r,d]$-linear code over $\F_q$ for any $1\le r\le k-1$.
\end{lemma}
\subsection{Algebraic geometry codes}
Let us introduce some background on algebraic curves over finite fields and their function fields. The reader may refer to \cite{St09,TV91,NX01} for details.

Let $F$ be a function field with the full constant field $\F_q$. In case there is no confusion, we simply use $g$ to denote  the genus $g(F)$.

Let $\PP_F$ be the set of  places. 
A place $P$ on $F$ is called rational  if the degree $P$ is $1$. A divisor $G$ of $F$ is a  formal sum $G=\sum_{P\in\PP_{F}}  n_PP$, where  $n_P$ are integers and almost all $n_P$ are equal to zero. $G$ is called {\it effective} if $n_P\ge 0$ for all $P\in\PP_{F}$. The degree $\deg(G)$ of $G$ is defined by $\deg(G)=\sum_{P\in\PP_{F}}  n_P\deg(P)$. 

The support of $G$, denoted by $\Supp(G)$, is defined to be the set $\{P\in\PP_{F}:\; n_P\not=0\}$. For a point $P$, we denote by $\nu_P$ the normalized discrete valuation of $P$. For a nonzero function $x$ in $F$, one can define a principal divisor $(x):=\sum_{Q\in\PP_{F}} \nu_Q(x)Q$. It is a well-known fact that the degree of $(x)$ is zero. The zero divisor and pole divisor of $x$ are defined to be $(x)_0:=\sum_{P\in\PP_{F},\nu_P(x)>0} \nu_P(x)P$ and $(x)_{\infty}:=-\sum_{P\in\PP_{F},\nu_P(x)<0} \nu_P(x)P$, respectively. 

For a  rational place $P$, there exists a function $t_P\in F$ such that $\nu_P(t_P)=1$. It is called a local parameter at $P$.  A local parameter is not unique. For a nonzero function $f\in F$ with $\nu_P(f)=v\in\ZZ$, let $a_v=(t_P^{-v}f)(P)\in\F_q$. Then we have $\nu_P(t_P^{-v}f-a_v)\ge 1$, i.e.,  $\nu_P(((t_P^{-v}f)-a_v)/t_P)\ge 0$. Put $((t_P^{-v}f-a_v)/t_P)(P)=a_{v+1}\in\F_q$, then we have $\nu_P((t_P^{-v}f-a_v)/t_P-a_{v+1})\ge 1$, i.e, $\nu_P((t_P^{-v}f-a_v-a_{v+1}t_P)/t_P^2)\ge 0$. Put $((t_P^{-v}f-a_v-a_{v+1}t_P)/t_P^2)(P)=a_{v+2}\in\F_q$. Continue in this fashion, we get a sequence $\{a_i\}_{i=v}^{\infty}$ of elements in $\F_q$ such that
\begin{equation}\label{eq:}
\nu_P\left({t_P^{-v}f-\sum_{i=v}^ma_it_P^{i-v}}\right)\ge m+1-v\end{equation}
for all $m\ge v$. We denote by $f^{(i)}(P)$ the coefficient $a_i$ for all $i\ge v$. It is clear that $f^{(0)}(P)=f(P)$ does not depend on choice of $t_P$. On the other hand, $f^{(i)}(P)$ depends on choice of $t_P$ for $i\neq0$. Note that if $P$ is rational, then all $a_i$ are elements of $\F_q$. The series $\sum_{i=v}^{\infty}a_it_P^{  i-v}$ is called a {\it local expansion} of $f$ at $P$.

Denote by $\GO_F$ the set of differentials of $F$. Then $\GO_F$  forms a one-dimensional space over $F$.  It is well-know that $dt_P$ is not $0$ for any place $P$, where $t_P$ is a local parameter at $P$. Thus, every differential can be written as $ydt_P$ for some $y\in F$.  For a differential $\Go\in\GO_F$ and a point $P$, write $\Go=fdt_P$, we define the evaluation of $\Go$ at $P$ by $\nu_P(f)$. The divisor $(\Go)$ is defined to be $\sum_{P\in \PP_F}\nu_P(\Go) P$.  It is called a canonical divisor and has degree $2g-2$. It is clear that all canonical divisors are equivalent. 
 
For a rational place $P$, assume that $f$ has a local expansion $\sum_{i=v}^{\infty}a_vt_P^i$, define the residue of $\Go$ at $P$ to be $a_{-1}$ if $v<0$; and $0$ if $v\ge 0$, denoted by $\res_P(\Go)$. It is easy to verify that $\res_P(\Go)$ is independent of choice of the local parameter $t_P$. 
 
For a rational divisor $G$, define $\GO_F(G)$ to be the set 

\begin{equation}\label{eq:} \{\Go\neq 0\in\GO_F:\; (\Go)\ge G\}\cup\{0\}.\end{equation} Then  $\GO_F(G)$ is a  finitely dimensional space over $\F_q$ and its dimension, denoted by $i(G)$.

Now for a rational divisor $G$, one can
define the Riemann-Roch space $\mL(G)$ associated with $G$ by
\begin{equation}\label{eq:} \mL(G):=\{x\in F^*:\; (x)+G\ge 0\}\cup\{0\}.
\end{equation}
Then  $\mL(G)$ is a finitely dimensional space over $\F_q$ and its dimension is denoted by $\ell(G)$. Then for any canonical divisor $W=(\Go)$, we have the following $\F_q$-isomorphism
\begin{equation}\label{eq:4} \mL(W-G)\rightarrow\GO_F(G);\quad x\mapsto x\Go.
\end{equation}

The Riemann-Roch theorem states that
\begin{equation}
\ell(G)= \deg(G)-g+1+\ell(W-G).
\end{equation}
Hence, we have $\ell(G)= \deg(G)-g+1$, and the equality holds if $\deg(G)\ge 2g-1$ as $\ell(W-G)=0$ in this case.

Now let $P_1,\dots,P_n$ be $n$ pairwise distinct rational points. Put $D=\sum_{i=1}^nP_i$. Let $G$ be a rational divisor with $\Supp(G)\cap\Supp(D)=\emptyset$. For an vector $\bu=(u_1,\dots,u_n)\in(\F_q^*)^n$,  define the following generalized functional code
\[\mC_L(D,G,\bu)=\{(u_1f(P_1),\dots,u_nf(P_n)):\; f\in\mL(G)\},\]
and the differential code
\[\mC_\GO(D,G,\bu)=\{(u_1\res_{P_1}(\Go),\dots,u_n\res_{P_n}(\Go)):\; \Go\in\GO(G-D)\}.\]
These two codes are called generalized Goppa geometric codes. 
The dual code of $\mC_L(D,G,{\bf 1})$ is the code $\mC_\GO(D,G,{\bf 1})$, where ${\bf 1}$  is the all-one vector. Furthermore, the $\F_q$-dimension of $\mC_\GO(D,G,\bu)$ is $\ell(G)$ if $0\le \deg(G)<n$. 

Assume \(0\le\deg G<n\) and \(\operatorname{Supp}(G)\cap\operatorname{Supp}(D)=\varnothing\). The kernel of evaluation on \(\mathcal L(G)\) is \(\mathcal L(G-D)=\{0\}\). Hence
\[
\dim C_L(D,G,\mathbf u)=\ell(G),
\qquad
d\bigl(C_L(D,G,\mathbf u)\bigr)\ge n-\deg G.
\]
Indeed, if a nonzero \(f\in\mathcal L(G)\) vanishes at \(s\) evaluation places, then the effective divisor \((f)+G\) has degree at least \(s\). In genus zero, \(\ell(G)=\deg G+1\). The Singleton bound therefore gives
\[
C_L(D,G,\mathbf u)\text{ is an }
[n,\deg G+1,n-\deg G]\text{ MDS code}.
\]
\subsection{Rational function fields and generalized Reed-Solomon codes}
In this section, we assume that $F$ is the rational function field $\F_q(x)$,  where $x$ is a transcendental element over $\F_q$. Hence, the genus of $F$ is $0$. Furthermore, every place uniquely  correspondences a monic irreducible polynomial or $1/x$.

Let $r, s\ge 1$ be integers. A rational place is either the zero of $x-\Ga$ for some $\Ga\in \F_q$ or the pole of $x$, i.e., the zero of $1/x$. Denote by $P_\Ga$  the zero of $x-\Ga$. We also denote by $\Pin$ the pole of $x$. Thus, if $G$ is a divisor \begin{equation}\label{eq:2.3.1}
    \sum_{i=1}^ra_iP_i-\sum_{j=1}^sb_jQ_j+m\Pin
    \end{equation}
    with $a_i,b_j>0$ and 
    $\deg(G)= \sum_{i=1}^ra_i\deg(P_i)-\sum_{j=1}^sb_j\deg(Q_j)+m\geq 0$, where $P_i$ and $Q_j$ correspondence to the monic irreducible polynomials $p_i(x)$ and $q_j(x)$, respectively. Then we can deduce from the definition that the Riemman-Roch space $\mL(G)$ is 
\[\mL(G)=\left\{\frac{\prod_{j=1}^sq_j(x)^{b_j}}{\prod_{i=1}^rp_i(x)^{a_i}}f(x):\; f(x)\in\F_q[x]_{\le deg(G)}\right\}. \]

If $G$ is a divisor of degree $k-1$ with $k\ge 1$, we have by the Riemann-Roch theorem that $\ell(G)=\deg(G)+1=k$. In particular, we have 
\[\mL((k-1)\Pin)=\{f(x)\in\F_q[x]:\; \deg(f)<k\}.\]

Let $\Ga_1,\dots,\Ga_n$ be pairwise distinct elements. Then
the functional code
\begin{eqnarray*}
\mC_L(D,(k-1)\Pin,\bu)&=&\{(u_1f(P_{\Ga_1}),\dots,u_nf(P_{\Ga_n})):\; f\in\mL((k-1)\Pin)\}\\
&=&\{(u_1f(\Ga_1),\dots,u_nf(\Ga_n)):\; f\in\F_q[x]_{<k}\}.
\end{eqnarray*}
is a generalized Reed-Solomon (GRS) code, where $D=\sum_{i=1}^nP_{\Ga_i}$ and $\bu\in(\F_q^*)^n$.

Let us now consider an arbitrary divisor $G$ given in \eqref{eq:2.3.1}.  
Let $\Ga_1,\dots,\Ga_n$ be pairwise distinct elements satisfying $\Supp(G)\cap\Supp(D)=\emptyset$, where $D=\sum_{i=1}^nP_{\Ga_i}$. Then for $\bu=(u_1,\dots,u_n)\in(\F_q^*)^n$, 
the functional code
\begin{eqnarray*}
\mC_L(D,G,\bu)&=&\{(u_1f(P_{\Ga_1}),\dots,u_nf(P_{\Ga_n})):\; f\in\mL(G)\}\\
&=&\{(v_1u_1h(\Ga_1),\dots,v_nu_nh(\Ga_n)):\; h\in\F_q[x]_{\le \deg(G)}\}.
\end{eqnarray*}
is a GRS code, where $v_i=\frac{\prod_{j=1}^sq_j(\Ga_i)^{b_j}}{\prod_{i=1}^rp_i(\Ga_i)^{a_i}}$.

Finally, we consider consider an arbitrary divisor $G$ with $m=0$ and $\deg(G)=k-1$ given in \eqref{eq:2.3.1}.  
Let $\Ga_1,\dots,\Ga_{n-1}$ be pairwise distinct elements satisfying $\Supp(G)\cap\Supp(D)=\emptyset$, where $D=\sum_{i=1}^{n-1}P_{\Ga_i}+P_n$ with $P_n=\Pin$. Then for $\bu=(u_1,\dots,u_n)\in(\F_q^*)^n$,
the functional code
\begin{eqnarray*}
\mC_L(D,G,\bu)&=&\{(u_1f(P_{\Ga_1}),\dots,u_{n-1}f(P_{\Ga_{n-1}}),u_nf(P_n)):\; f\in\mL(G)\}\\
&=&\{(u_1v_1h(P_{\Ga_1}),\dots,v_{n-1}u_{n-1}h(P_{\Ga_{n-1}}),u_nh_{k-1}):\; h\in\F_q[x]_{<k}\}.
\end{eqnarray*}
is an extended GRS code. Note that $h(\Pin)=f_{k-1}$, where $h_{k-1}$ is the  coefficient of $x^{k-1}$ of $h(x)$  and $v_i=\frac{\prod_{j=1}^sq_j(\Ga_i)^{b_j}}{\prod_{i=1}^rp_i(\Ga_i)^{a_i}}$.

For later use in the proof, we state the following results.

\begin{lemma}\label{lem:4.1}
    For a nonzero  function $f$ in $F$, the divisor $(fdx)$ is equal to $(f)-2\Pin$.
\end{lemma}
\begin{proof}
    For any place $P$ corresponding to a monic irreducible polynomial $p(x)$,  $p(x)$ is a local parameter at $P$. Thus, we have 
\[\nu_{P}(fdx)=\nu_{P}\left(\frac{f(x)}{p'(x)}d(p(x))\right)=\nu_{P}\left(\frac{f(x)}{p'(x)}\right)=\nu_P(f).\] As for $\Pin$, $\frac1x$ is a local parameter. Thus, we have
    \[\nu_{\Pin}(fdx)=\nu_{\Pin}\left(-x^2fd\left(\frac1x\right)\right)=\nu_{\Pin}(f)-2\Pin.\] The desired result follows. 
\end{proof}

In \(\mathbb F_q(x)\), one has \((dx)=-2P_\infty\) and \((R\,dx)=(R)-2P_\infty\) for every nonzero rational function \(R\). If \(H\in\mathbb F_q[x]\) has a simple root \(\alpha\) and \(A\) is regular at \(P_\alpha\), then
\[
\operatorname{res}_{P_\alpha}\!\left(\frac{A(x)}{H(x)}\,dx\right)
=\frac{A(\alpha)}{H'(\alpha)}.
\]
This follows by writing \(H(x)=(x-\alpha)H_\alpha(x)\), where \(H_\alpha(\alpha)=H'(\alpha)\). At \(P_\infty\), use \(t=1/x\) and \(dx=-t^{-2}dt\). In particular, if the leading term of \(R(x)\) at infinity is \(c/x\), then \(R(x)\,dx\) has a simple pole at \(P_\infty\) with residue \(-c\).

More generally, if \(\eta\) has a simple pole at \(P\) and \(f\) is regular there, then \(\operatorname{res}_{P}(f\eta)=f(P)\operatorname{res}_{P}(\eta)\).

\subsection{Finite-field polynomial identities}
Let \(q=p^e\). If \(V\subseteq\mathbb F_q\) is an \(\mathbb F_p\)-subspace of dimension \(r\), then its subspace polynomial
\[
F_V(X)=\prod_{\alpha\in V}(X-\alpha)
\]
is \(p\)-linearized, meaning that its only possible nonzero terms have exponents \(p^j\). In particular, \(F_V'(X)=\beta_V\), where
\[
\beta_V=[X]F_V(X)=\prod_{\alpha\in V\setminus\{0\}}(-\alpha)\ne0.
\]
To see the linearized form, start with \(F_{\{0\}}(X)=X\). If \(V=U\oplus\mathbb F_p b\), then
\[
F_V(X)=F_U(X)^p-F_U(b)^{p-1}F_U(X).
\]
For odd \(p\), pairing the nonzero elements of \(V\) as \(\{\alpha,-\alpha\}\) gives
\[
\beta_V=(-1)^{(|V|-1)/2}
\prod_{\{\alpha,-\alpha\}\subset V\setminus\{0\}}\alpha^2,
\]
where the product runs over the distinct pairs. Thus \(\beta_V\) is a square whenever \(q\equiv1\pmod4\).

If \(N\mid q-1\), the subgroup of \(\mathbb F_q^\times\) of order \(N\) has vanishing polynomial \(X^N-1\), with derivative \(NX^{N-1}\). Also, the squaring map is bijective in characteristic two. For odd \(q\), \(-1\) is a square precisely when \(q\equiv1\pmod4\).

\subsubsection{Projective arcs and hyperovals}
For \(k\ge2\), the columns of a generator matrix of a linear \([n,k]\) MDS code represent an \(n\)-arc in \(\operatorname{PG}(k-1,q)\), meaning that no \(k\) of these points lie in a hyperplane. Conversely, choosing a nonzero representative of each point of an \(n\)-arc gives the columns of an MDS generator matrix. The rational normal curve consists of the points represented by
\[
(1,a,\ldots,a^{k-1})^{\mathsf T},\qquad a\in\mathbb F_q,
\]
together with \((0,\ldots,0,1)^{\mathsf T}\). An arc projectively equivalent to this curve yields an extended GRS code, with the choice of representatives determining the column multipliers.

For even \(q\), a \((q+2)\)-arc in \(\operatorname{PG}(2,q)\) is a hyperoval. A hyperoval is regular if it is the union of a nonsingular conic and its nucleus. The nucleus is the common point of the tangents to the conic. Every other point off the conic lies on a tangent. Consequently, a hyperoval containing the \(q+1\) points of a nonsingular conic must be regular. Every hyperoval is regular for \(q=4,8\), and a nonregular hyperoval exists for every \(q=2^m\) with \(m\ge4\).

\section{A framework of construction}
In this section, we establish the framework of our construction scheme.
\begin{lemma}~\label{lem:3.1} For a vector $\bu\in(\F_q^*)^n$, 
the dual code $\mC_L^{\perp}(D,G,\bu)$ of $\mC_L(D,G,\bu)$ is $\mC_\GO(D,G,\bu^{-1})$, where where $\bu^{-1}=(u_1^{-1},\dots,u_n^{-1})$.
\end{lemma}
\begin{proof}
This follows from the fact that     the dual code $\mC_L^{\perp}(D,G,{\bf 1})$ of $\mC_L(D,G,{\bf 1})$ is $\mC_\GO(D,G,{\bf 1})$.
\end{proof}

\begin{lemma}~\label{lem:3.2} If $\eta$ is a differential satisfying $\nu_{P_i}(\eta)=-1$ for all $1\le i\le n$, then the  dual code $\mC_L^{\perp}(D,G,\bu)$ of $\mC_L(D,G,\bu)$ is $\mC_L(D, D-G+(\eta), \bu^{-1}*{\bf res}(\eta))$, where ${\bf res}(\eta)=(\res_{P_1}(\eta), \dots,\res_{P_n}(\eta))$.
\end{lemma}
\begin{proof} By Lemma~\ref{lem:3.1}, it is sufficient to show that $\mC_L(D, D-G+(\eta), \bu^{-1}*{\bf res}(\eta))$ is equal to $\mC_\GO(D,G,\bu^{-1})$. As $\nu_{P_i}(\eta)=-1$ for all $1\le i\le n$, we have $\supp(D)\cap\supp(D-G+(\eta))=\emptyset$. Thus, the code $\mC_L(D, D-G+(\eta), \bu^{-1}*{\bf res}(\eta))$ is well defined. For any $(h(P_1)u_1^{-1}\res_{P_1}(\eta),\dots,h(P_n)u_n^{-1}\res_{P_n}(\eta))\in \mC_L(D, D-G+(\eta), \bu^{-1}*{\bf res}(\eta))$ with $h\in\mL(D-G+(\eta))$, we have
\[(h(P_1)u_1^{-1}\res_{P_1}(\eta),\dots,h(P_n)u_n^{-1}\res_{P_n})=(u_1^{-1}\res_{P_1}(h\eta),\dots,u_n^{-1}\res_{P_n}(h\eta))\in\mC_\GO(D,G,\bu^{-1}). \]
The inclusion in the above display follows from fact that $h\eta\in \GO(G-D)$. This implies that $\mC_L(D, D-G+(\eta), \bu^{-1}*{\bf res}(\eta))\subseteq \mC_\GO(D,G,\bu^{-1})$.
On the other hand, by the isomorphism given in \eqref{eq:4},  every element of $\GO(G-D)$ has the form $h\eta$ for some $h\in \mL(D-G+(\eta))$. Hence, we must have $\mC_L(D, D-G+(\eta), \bu^{-1}*{\bf res}(\eta))=\mC_\GO(D,G,\bu^{-1})$.
\end{proof}

\begin{theorem}\label{thm:3.3}
If there exists a differential $\eta$ and a vector $\bu=(u_1,\dots,u_n)\in(\F_q^*)^n$ satisfying 
\begin{itemize}
\item[{\rm (i)}]
 $\nu_{P_i}(\eta)=-1$ for all $1\le i\le n$; 
\item[{\rm (ii)}] $\bu^2={\bf res}(\eta)$, where $\bu^2=(u_1^2,\dots,u_n^2)$; and 
\item[{\rm (iii)}] $D+(\eta)\ge 2G$, 
\end{itemize}
then the code $\mC_L(D,G,\bu)$ is self-orthogonal. 
\end{theorem}
\begin{proof} As $\bu^2={\bf res}(\eta)$, we have $\bu^{-1}*{\bf res}(\eta)=\bu$. Thus, we have $\mC_L(D, D-G+(\eta), \bu^{-1}*{\bf res}(\eta))=\mC_L(D, D-G+(\eta), \bu)$. Since $D+(\eta)\ge 2G$, we further have $\mC_L(D,G,\bu)\subseteq \mC_L(D, D-G+(\eta), \bu)$. This gives $\mC_L(D,G,\bu)\subseteq \mC_L(D, D-G+(\eta), \bu^{-1}*{\bf res}(\eta))$. Now the desired result follows from Lemma~\ref{lem:3.2}  showing that the dual of  $\mC_L(D,G,\bu)$  is  $\mC_L(D, D-G+(\eta), \bu^{-1}*{\bf res}(\eta))$.  
\end{proof}
\section{Instantiation}
In this section, we study the case of rational function field, i.e,  $F=\F_q(x)$  and conduct discussions separately for even and odd values of $q$.

\subsection{$q$ is even}
In this subsection, we consider  $q$-ary self-orthogonal MDS codes when $q$ is even.
\begin{theorem}\label{thm:4.2}
   For any $n\le q$ and $k\le n/2$, there exists a $q$-ary $[n,k]$ self-orthogonal MDS code. 
\end{theorem}
\begin{proof}
  Let $\Ga_1,\dots,\Ga_n$  be pairwise distinct elements of $\F_q$. Let $P_{i}$ be the zero of $x-\Ga_i$. Let $D$ be the divisor  $\sum_{i=1}^nP_{i}$. Let $G$ be the divisor $(k-1)\Pin$. Consider the differential $\eta=\frac{dx}{\prod_{i=1}^n(x-\Ga_i)}$. Then $(\eta)=-D+n\Pin-2\Pin=-D+(n-2)\Pin$. Thus,  $\nu_{P_i}(\eta)=-1$ and $D-G+(\eta)=D-(k-1)\Pin+(n-2)\Pin-D=(n-k-1)\Pin\ge (k-2)\Pin=G$
  {as $k\le n/2.$} 
  As every element of $\F_q$ is a square, there exists a vector $\bu$ satisfying $\bu^2={\bf res}(\eta)$. Since $\nu_{P_i}(\eta)=-1$, $\res_{P_i}(\eta)$ is nonzero. Hence, $\bu$ is a vector in $(\F_q^*)^n$. The desired result follows from Theorem.
\end{proof}
\begin{rem}{\rm
    The above code can be realized via the elementary construction. Let $\Ga_1,\dots,\Ga_n$ be pairwise distinct elements of $\F_q$ with $n\le q$. Consider the GRS code $\RS_k(\Ga, \bu)$. Its dual code is $\RS_{n-k}(\Ga, \bv)$, where $\bv=(v_1,\dots,v_n)$ is a nonzero solution of the equation 
\begin{equation}\left(\begin{matrix}
    1&1&\cdots&1\\
     \Ga_1&\Ga_2&\cdots&\Ga_n\\
\Ga_1^2&\Ga_2^2&\cdots&\Ga_n^2\\
        \cdots&\cdots&\cdots&\cdots\\
        \Ga_1^{n-2}&\Ga_2^{n-2}&\cdots&\Ga_n^{n-2}
\end{matrix}\right)\left(\begin{matrix}
    u_1x_1\\
    u_2x_2\\ 
      u_3x_3\\ 
    \vdots\\
     u_nx_n
\end{matrix}\right)=\bo.
\end{equation}
If $k\le n/2$ and $\bv=\bu$, i.e., $\bu^2$ is a solution of \begin{equation}\label{eq:8}
\left(\begin{matrix}
    1&1&\cdots&1\\
     \Ga_1&\Ga_2&\cdots&\Ga_n\\
\Ga_1^2&\Ga_2^2&\cdots&\Ga_n^2\\
        \cdots&\cdots&\cdots&\cdots\\
        \Ga_1^{n-2}&\Ga_2^{n-2}&\cdots&\Ga_n^{n-2}
\end{matrix}\right)\left(\begin{matrix}
    y_1\\
    y_2\\ 
      y_3\\ 
    \vdots\\
     y_n
\end{matrix}\right)=\bo,
\end{equation}
then $\RS_k^{\perp}(\Ga, \bu)=\RS_{n-k}(\Ga, \bv)=\RS_{n-k}(\Ga, \bu)\supseteq \RS_{k}(\Ga, \bu)$, i.e., $\RS_{k}(\Ga, \bu)$ is self-orthogonal. To see that the code $\RS_{k}(\Ga, \bu)$ coincides with the code $\mC_L(D,(k-1)\Pin, \bu)$ given in Theorem \ref{thm:4.2}, we need to show that ${\bf res}(\eta)$ is a solution of  \eqref{eq:8}. Note that the solution of \eqref{eq:8} is $\left(\frac{1}{\prod_{j\neq 1}(\Ga_1-\Ga_j)},\dots,\frac{1}{\prod_{j\neq n}(\Ga_n-\Ga_j)}\right) $.

Since $\eta=\frac1{x-\Ga_i}\times \frac{1}{\prod_{j\neq i}(x-\Ga_j)}d(x-\Ga_j)$, we have $\res_{P_i}(\eta)=\frac{1}{\prod_{j\neq i}(\Ga_i-\Ga_j)}$. 
}\end{rem}

Then, we investigate a self-orthogonal MDS code when the length $n=q+1$.
\begin{theorem}~\label{thm:4.4}
   For any $1\le k\le q/2-1$, there exists a $q$-ary $[q+1,k]$ self-orthogonal MDS code. 
\end{theorem}
\begin{proof} Let $\Ga_1,\dots,\Ga_{q}$  be pairwise distinct elements of $\F_q$. Let $P_{i}$ be the zero of $x-\Ga_i$ for $1\le i\le q$. Let $P_n$ be the pole $\Pin$ of $x$. Let $D$ be the divisor  $\sum_{i=1}^nP_{i}$.
Let $a(x)\in\F_q[x] $ be an irreducible polynomial of degree $\frac{q}2+2\ge 3$ and let $b(x)\in\F_q[x]$ be an irreducible polynomial of degree $\frac{q}2+2-(k-1)\ge 3$. Let $c(x)$ be an irreducible polynomial of degree $(q-1)-2(k-1)\ge 3$. Consider the differential $\eta=\frac{a^2(x)c(x)}{b^2(x)(x^q-x)}dx$. Then $(\eta)=2A+C-2B-\sum_{i=1}^{q}P_i+(2\deg(b(x))-2\deg(a(x)-\deg(c(x)))\Pin+(q-2)\Pin=2A+C-2B-D$, where $A,B,C$ are the divisors satisfying $(a(x))=A-\deg(a(x))\Pin$, $(b(x))=B-\deg(b(x))\Pin$ and $(c(x))=C-\deg(c(x))\Pin$. { Put $G=A-B$, then $\deg(G)=k-1$.} Thus, $\mC_L(D,G)$ is a $q$-ary $[n,k]$-MDS code. As $\nu_{P_i}(\eta)=-1$ and $D-G+(\eta)=D-G+2A+C-2B-D=A-B+C\ge A-B=G$. 
  As every element of $\F_q$ is a square, there exists a vector $\bu$ satisfying $\bu^2={\bf res}(\eta)$. Since $\nu_{P_i}(\eta)=-1$, $\res_{P_i}(\eta)$ is nonzero. Hence, $\bu$ is a vector in $(\F_q^*)^n$. The desired result follows from Theorem \ref{thm:3.3}.
\end{proof}
The above two theorems completely solves existence of $q$-ary self-orthogonal MDS codes for codes of length up to $q+1$ except for the case where code length is $q+1$ and dimension is $q/2$. The following result shows nonexistence of  $[q+1,q/2]$-orthogonal MDS codes under the conjecture of the Main MDS conjecture. Before proving this nonexistence for arbitrary $q$, let us show that it is true for $q=2$ and $4$.
\begin{prop}\label{prop:4.5}
   There are no $q$-ary $[q+1,q/2]$ self-orthogonal MDS codes for $q=2$ and $4$. 
\end{prop}
\begin{proof}
If $q=2$, a binary  $[q+1,q/2]$ MDS code is a binary $[3,1]$ MDS code. Thus, the distance of this code is $3$ and hence the code must be $\{(000),(111)\}$. However, this code is not self-orthogonal.

Then, we focus on $q=4$. Suppose that there were a $4$-ary $[5,2,4]$ self-orthogonal MDS code $\mC$. Assume that two vectors $(1,0,\Ga_1,\Ga_2,\Ga_3)$ and $(0,1,\Gb_1,\Gb_2,\Gb_3)$ form a systematic generator matrix $G$ of $\mC$ such that
\[\begin{bmatrix}
1 & 0 & \Ga_1 & \Ga_2 &\Ga_3 \\
0 & 1 &\Gb_1 &\Gb_2 &\Gb_3
\end{bmatrix}\]
Thus, $\Ga_i\neq 0$ and $\Gb_i\neq 0$ for all $1\le i\le 3$. Since the inner product of $(1,0,\Ga_1,\Ga_2,\Ga_3)$ with itself is $0$, we get $1+\Ga_1^2+\Ga_2^2+\Ga_3^2=0$, i.e., $\Ga_1+\Ga_2+\Ga_3=1$. Similarly, $\Gb_1+\Gb_2+\Gb_3=1$. This also implies that the all-one vector $(11111)$ belongs to $\mC^\perp$. Since $1+\Ga+(\Ga+1)=0$, it is clear that $\{\Ga_1,\Ga_2,\Ga_3\}=\{1,\Ga,\Ga\}$, where $\Ga$ is a root of $x^2+x+1$. This implies that two of $\Ga_1,\Ga_2,\Ga_3$ must be equal. Without loss of generality, we may assume that $\Ga_1=\Ga_2$, then we must have $\Ga_3=1$ as $\Ga_1+\Ga_2+\Ga_3=1$. Hence, $(1,0,\Ga_1,\Ga_2,\Ga_3)=(1,0,a,a,1)$. After scaling, the codeword $(0,1,\Gb_1,\Gb_2,\Gb_3)$ becomes $(0,\Gb_3^{-1},u,v,1)$ for some nonzero elements $u,v\in\F_4$. Thus, we have $ua+va+1=0$, i.e., $u+v=a^{-1}$. As two of $u,v,1$ are equal, if $u=v$  the codeword  \[(1,0,a,a,1)-u^{-1}a(0,\Gb_3^{-1},u,u,1)=(1,0,a,a,1)-(0, u^{-1}a\Gb_3^{-1}, a, a,u^{-1}a)\] would be Hamming weight at most $3$. If  $u\neq v$, we may assume that $u=1$ and hence $v=1+a^{-1}=a$. This gives a contradiction as the codeword \[(1,0,a,a,1)-(0,\Gb_3^{-1},1,a,1)\] has the Hamming weight at most $3$. This completes the proof.
\end{proof}

\begin{theorem}~\label{thm:4.6}
  If the Main MDS holds, then  there are no $q$-ary $[q+1,q/2,q/2+2]$ self-orthogonal MDS codes for even $q$.
\end{theorem}
\begin{proof} By Proposition \ref{prop:4.5}, we can assume that $q\ge 8$.

Suppose there were a $q$-ary $[q+1,q/2]$ self-orthogonal MDS code $\mC$. Then $\mC^\perp$ is a $q$-ary $[q+1,q/2+1]$  MDS code. Choose a vector $\bc\in\mC^\perp\setminus \mC $. Let $\mC_1$ be the $q$-ary  $[q+2,q/2, q/2+2]$ code obtaining from $\mC$ by adding $0$ to every codeword of $\mC$ such that 
\[\mC_1=\{(\bu, 0):\bu\in\mC\}\subseteq\F_q^{q+2},\] then we have $\dim(\mC_1)=\dim(\mC)=q/2$. Let $\mA$ be the code spanned by $\mC_1$ and $\bv=(\bc,1)$ such that
\[\mA=\rm{Span}\{\mC_1, \bv\}.\]We claim that $\bv\notin\mC_1$. Suppose that $\bv\in\mC_1$,  there exits $\bu\in \mC$ such that $(\bc,1)=(\bu,0), 1=0\in\F_q$. There is a contradiction. Hence, $\bv\notin \mC_1$. The dimension of $\mA$ is $\dim(\mA)=\dim(\mC_1)+1=q/2+1$. It is clear that the code $\mA$ has length $q+2$. We claim that the minimum distance of $\mA$ is at least $q/2+2$, i.e., $\mA$ is a $q$-ary $[q+2,q/2+1,q/2+2]$-MDS code. Indeed, assume that $\bu'$ is a nonzero codeword of $\mA$. If $\bu'\in\mC_1$, then $\wt(\bu')\ge q/2+2$ as $\mC_1$ has the minimum distance $q/2+2$. If $\bu'\not\in\mC_1$, then $\bu'=\Ga(\bc_1,0)+\Gb(\bc,1)$ for some $\bc_1\in \mC$, $\Ga\in\F_q$ and $\Gb\in\F_q^*$. As $\Ga\bc_1+\Gb\bc$ is a nonzero codeword of $\mC^\perp$, it has weight at least $q/2+1$. Hence, $\bu'=\Ga(\bc_1,0)+\Gb(\bc,1)$ has weight at least $q/2+2$. 

In case $q\ge8$, for dimension of $\mA$ we have $q/2+1\neq 3$ and $q-1$, hence there are no $q$-ary $[q+2,q/2+1,q/2+2]$-MDS codes under the Main MDS conjecture. By Lemma~\ref{lem.P1}, we have there are no $q$-ary $[q+1,q/2,q/2+2]$-MDS codes under  the Main MDS conjecture.
\end{proof}
 By the Main MDS conjecture, we have $q$-ary $[q+2,3,q-1]$-MDS codes. The following Proposition \ref{prop:4.7x} and Theorem \ref{thm:4.7} completely solve self-orthogonality for $q$-ary $[q+2,3,q-1]$-MDS codes.
\begin{prop}~\label{prop:4.7x}
Let $q$ be a power of $2$ with $4\le q<16$. There is no self-orthogonal $[q+2,3]_q$ MDS code.
\end {prop}
Proof. Suppose $\mC$ is such a code, and let \(G\) be a generator matrix. Since $\mC$ is MDS, any three columns of \(G\) are linearly independent. Thus the columns represent a set \(S\) of \(q+2\) distinct points in \(\mathrm{PG}(2,q)\) with no three collinear.

Let $q$ is a power of \(2\). The stated range gives \(q=4\) or \(q=8\). In both cases every hyperoval in \(\mathrm{PG}(2,q)\) is regular. Thus \(S\) consists of a nonsingular conic and its nucleus. After an invertible row operation on \(G\), we may take the conic to be \(XZ+Y^2=0\) and its nucleus to be \((0:1:0)\). Such a row operation preserves the condition \(GG^{\mathsf T}=0\). 

Writing the resulting columns as \((x_i,y_i,z_i)^{\mathsf T}\), self-orthogonality gives
\[
0=(GG^{\mathsf T})_{13}+(GG^{\mathsf T})_{22}
 =\sum_{i=1}^{q+2}(x_i z_i+y_i^2),
\]
where $(GG^{\mathsf T})_{ij}$ stands for the inner product of the $i$th and $j$th rows. 
All conic columns contribute zero. The nucleus column is \((0,\lambda,0)^{\mathsf T}\) for some \(\lambda\in\mathbb F_q^\times\), so the sum equals \(\lambda^2\ne0\). This is a contradiction. \(\square\)

We now turn to the case where $q\ge 16$.

{\begin{theorem}~\label{thm:4.7}
 Let \(q\) be a power of two. A linear Euclidean self-orthogonal \([q+2,3,q-1]_q\) MDS code exists if and only if \(q\ge16\).

More precisely, for \(q\ge8\), a hyperoval in \(\mathrm{PG}(2,q)\) admits a choice of column representatives that generate a Euclidean self-orthogonal MDS code if and only if the hyperoval is nonregular.
\end{theorem}
\begin{proof}
The columns of a generator matrix of a \([q+2,3]_q\) MDS code form a hyperoval. For a regular hyperoval, choose projective coordinates in which the conic is \(XZ+Y^2=0\) and the nucleus is \((0:1:0)\). If the columns are \((x_i,y_i,z_i)^{\mathsf T}\), self-orthogonality implies
\[
0=(GG^{\mathsf T})_{13}+(GG^{\mathsf T})_{22}
=\sum_i(x_i z_i+y_i^2).
\]
Every conic column contributes zero, while the nucleus column contributes a nonzero square. This rules out regular hyperovals for any choice of column representatives.

Conversely, let \(v_1,\ldots,v_{q+2}\) represent a nonregular hyperoval and set
\[
K=\ker\!\left(w\longmapsto \sum_{i=1}^{q+2}w_i v_i v_i^{\mathsf T}\right).
\]
The codimension of \(K\) is at most six. Every coordinate functional is nonzero on \(K\). If coordinate \(i\) vanished identically on \(K\), the row space of the displayed map would contain the \(i\)-th coordinate functional. This would give a quadratic vanishing at the other \(q+1\) hyperoval points. A singular quadratic contains at most four hyperoval points. A nonsingular conic containing \(q+1\) of the points forces the remaining point to be its nucleus, since any other point off the conic lies on a tangent and would produce three collinear hyperoval points. This contradicts nonregularity.

Choose a basis of \(K^*\) from the coordinate functionals. Assign its coordinates independently and uniformly in \(\mathbb F_q^\times\). Each of the at most six remaining coordinates vanishes with probability at most \(1/(q-1)\), since it is a nonzero linear combination of the chosen coordinates. For \(q\ge8\), the union bound gives \(6/(q-1)<1\). Hence there is a vector \(w\in K\) with every entry nonzero. Taking the columns to be \(\sqrt{w_i}\,v_i\) gives \(GG^{\mathsf T}=0\) and preserves the MDS property.

Finally, every hyperoval is regular for \(q=4,8\), and nonregular hyperovals exist for every \(q=2^m\) with \(m\ge4\), starting with the Lunelli–Sce hyperoval over \(\mathbb F_{16}\). For \(q=2\), self-orthogonality is impossible because \(2\cdot3>4\).
\end{proof}
}

\subsection{$q$ is odd}
In this subsection, we investigate the existence of certain $q$-ary self-orthogonal MDS codes for odd prime power $q$.

\begin{theorem}~\label{thm:4.8}  For any odd $q\ge 3$ and  
    any $1\le k\le (q+1)/2$ with $k\neq (q-1)/2$,  there exists a $q$-ary $[q+1,k]$ self-orthogonal MDS code. 
\end{theorem}

\begin{proof} 
For $q=3$, we need to show existence of a $[4,2]_3$ self-orthogonal MDS code. Indeed, the code generated by $(1011)$ and $(0112)$ is  a $[4,2]_3$ self-orthogonal MDS code.

Now assume that $q\ge 5$. 
Let $\Ga_1,\dots,\Ga_{q}$  be pairwise distinct elements of $\F_q$. Let $P_{i}$ be the zero of $x-\Ga_i$ for $1\le i\le q$. Let $P_n$ be the pole $\Pin$ of $x$. 
Let $a(x)\in\F_q[x] $ be a monic  irreducible polynomial of degree $k+1\ge 2$ and let $b(x)\in\F_q[x]$ be a monic  irreducible polynomial of degree $2$. If $k=(q+1)/2$, set $c(x)=1$. If $k\le (q-1)/2-1$, choose $c(x)$ to be a monic irreducible polynomial of degree $(q+1)/2-k\ge 2$. Consider the differential $\eta=-\frac{a^2(x)c^2(x)}{b^2(x)(x^q-x)}dx$. Then, we have
{\small \begin{eqnarray}~\label{eq:4.7.1}
\nonumber(\eta)&=&2A+2C-2B-\sum_{i=1}^{q}P_i+(2\deg(b(x))-2\deg(a(x)-2\deg(c(x)))\Pin+(q-2)\Pin\\
&=&2A+2C-2B-D\end{eqnarray}}
where $A,B,C$ are the divisors satisfying $(a(x))=A-\deg(a(x))\Pin$, $(b(x))=B-\deg(b(x))\Pin$ and $(c(x))=C-\deg(c(x))\Pin$. Let $D$ be the divisor  $\sum_{i=1}^nP_{i}$. Put $G=A-B$, then $\deg(G)=k-1$. Thus, $\mC_L(D,G)$ is a $q$-ary $[n,k]$-MDS code. As $\nu_{P_i}(\eta)=-1$ and $D-G+(\eta)=D-G+2A+2C-2B-D=A-B+2C\ge A-B=G$ from the equation~(\ref{eq:4.7.1}).

For each $1\le i\le q=n-1$, let $f_i(x):=\prod_{j\neq i}(x-\Ga_j)$. Then $(x-\Ga_i)f_i(x)=x^q-x$. Taking the derivatives for the both sides gives $f_i(x)+(x-\Ga_i)f_i'(x)=-1$. This gives $f_i(\Ga_i)=\prod_{j\neq i}(\Ga_i-\Ga_j)=-1$.
As $\eta=-\frac1{x-\Ga_i}\times \frac{a^2(x)c^2(x)}{b^2(x)\prod_{j\neq i}(x-\Ga_j)}d(x-\Ga_i)$, we have 
\[\res_{P_i}(\eta)=-\frac{a^2(\Ga_i)c^2(\Ga_i)}{b^2(\Ga_i)\prod_{j\neq i}(\Ga_i-\Ga_j)}=-\frac{a^2(\Ga_i)c^2(\Ga_i)}{b^2(\Ga_i)(-1)}=\frac{a^2(\Ga_i)c^2(\Ga_i)}{b^2(\Ga_i)}.\] Hence, $\res_{P_i}(\eta)$ is a nonzero square in $\F_q$ for all $1\le i\le q=n-1$.  

Let $t=\frac1x$. Then $\eta=x^2\times \frac{a^2(x)c^2(x)}{b^2(x)(x^q-x)}d\left(\frac{1}{x}\right)=\frac1t\times \frac{a^2(x)c^2(x)}{b^2(x)(x^{q-1}-1)}dt$. Thus, we have {$\res_{P_i}(\eta)=1$} which is a square. In conclusion, ${\bf res}(\eta)$ is equal to $\bu^2$ for some $\bu\in(\F_q^*)^n$.
 The desired result follows from Theorem \ref{thm:3.3}.
\end{proof}

 As in the even characteristic case, the above theorem does not yield a 
$[q+1,(q-1)/2]_q$
  self-orthogonal MDS code.

Let us show that self-dual MDS codes do not always produce self-orthogonal subcodes that are MDS. 
\begin{prop}~\label{prop:4.10}
   Assume the Main MDS conjecture. Then there are no $q$-ary $[q+1,(q-1)/2]$ self-orthogonal MDS codes contained in a $q$-ary $[q+1,(q+1)/2]$ self-dual MDS code.
\end{prop}
\begin{proof}
Let $\mC$ be a  $q$-ary $[q+1,(q+1)/2]$ self-dual MDS code. Suppose there were a subcode  $\mA$ of $\mC$  that is a $q$-ary $[q+1,(q-1)/2]$ self-orthogonal MDS code. Let $\{\bc_1,\dots,\bc_{(q-1)/2}\}$ be an $\F_q$-basis of $\mA$. Extend this basis to an $\F_q$ basis $\{\bc_1,\dots,\bc_{(q-1)/2}, $ $\bc_{(q+1)/2}\}$ of $\mC$. Then the code generated by $\{(0,\bc_1),\dots,(0,\bc_{(q-1)/2}),(1,\bc_{(q+1)/2})\}$ is a $q$-ary $[q+2,(q+1)/2]$-MDS code. This is contradictory to the Main MDS conjecture. 
\end{proof}

\begin{theorem}\label{thm:12x}
Let \(q\) be an odd prime power, then no extended generalized Reed–Solomon code with parameters \([q+1,(q-1)/2]_q\) is self-orthogonal. Furthermore,  for a prime \(q\)  or \(q=9\), no linear  self-orthogonal \([q+1,k]_q\) MDS codes exist.
\end{theorem}
\begin{proof}
First leet us consider an extended GRS code. Since its length is \(q+1\), its evaluation set is the entire projective line \(\mathbb F_q\cup\{\infty\}\). Its codewords have the form
\[
c(f)=\left((v_a f(a))_{a\in\mathbb F_q},\,
v_\infty[X^{k-1}]f\right),
\qquad \deg f<k,
\]
where every multiplier is nonzero. Set \(u_a=v_a^2\) and \(u_\infty=v_\infty^2\).

For each \(0\le j\le 2k-2=q-3\), choose \(r,s\in\{0,\ldots,k-1\}\) with \(r+s=j\). At infinity, the product of the coordinates of \(c(X^r)\) and \(c(X^s)\) is zero unless \(r=s=k-1\), in which case it is \(u_\infty\). Thus self-orthogonality requires
\begin{equation}\label{x1}
\sum_{a\in\mathbb F_q}u_a a^j+
u_\infty\,\mathbf 1_{\{j=q-3\}}=0,
\qquad 0\le j\le q-3.
\end{equation}
The \(q-2\) equations in \eqref{x1} are independent. Any linear dependence, restricted to the finite coordinates, would yield a polynomial of degree at most \(q-3\) vanishing at all \(q\) field elements. Hence the solution space has dimension \(3\).

For every \(h\in\mathbb F_q[X]\) with \(\deg h\le2\), the vector
\begin{equation}\label{x2}
u_a=h(a)\quad(a\in\mathbb F_q),\qquad
u_\infty=[X^2]h
\end{equation}
satisfies \eqref{x1}. Indeed, for \(j<q-3\), every monomial of \(X^jh(X)\) has degree less than \(q-1\), so its sum over \(\mathbb F_q\) is zero. For \(j=q-3\), the coefficient of \(X^{q-1}\) is \([X^2]h\), and \(\sum_{a\in\mathbb F_q}a^{q-1}=-1\). The map in \eqref{x2} is injective because \(q\ge3\), so \eqref{x2}  describes the entire solution space.

Since \(u_\infty\ne0\), the polynomial \(h\) is quadratic. Completing the square gives \(h(X)=A(X-B)^2+D\) with \(A\ne0\). As \(q\) is odd, \(h\) takes exactly \((q+1)/2\) distinct values on \(\mathbb F_q\). Yet each \(h(a)=v_a^2\) must be a nonzero square, and there are only \((q-1)/2\) nonzero squares. This contradiction proves the extended GRS claim.

For prime \(q\ge5\), Ball’s classification (see \cite{B12}) implies that the projective columns of a generator matrix of any linear \([q+1,k]_q\) MDS code form a normal rational curve. For \(q=9\), we have \(k=4\), and the corresponding \(10\)-arc in \(\mathrm{PG}(3,9)\) is a twisted cubic. In both cases, a change of row basis and a coordinate permutation put the generator matrix into the extended GRS form with arbitrary nonzero column multipliers. These operations preserve self-orthogonality, and the preceding argument covers every choice of multipliers.

Finally, for \(q=3\), an MDS generator of a \([4,1]_3\) code has four nonzero entries. Each entry has square \(1\), so the generator has self-inner product \(4=1\) in \(\mathbb F_3\). This completes the proof.
\end{proof}

In the following we show that if $q$ is divisible by the code length $n$ and the dimension satisfies $k\le n/2$, then there exists a self-orthogonal MDS code. 

\begin{theorem}~\label{thm:4.10}  Let $n|q$, then for
    any $1\le k\le n/2$,  there exists a $q$-ary $[n,k]$ self-orthogonal MDS code. 
\end{theorem}

\begin{proof} Let $q=p^\ell$, where $p$ is a prime and $\ell$ is an integer with $\ell\ge 1$. Choose an $\F_p$-subspace $V$ of $\F_q$ with $|V|=n$ and label the elements of $V$ by $V=\{\Ga_1,\dots,\Ga_n\}$. Then, we have the $\F_p$-linearized polynomial $f_V(x):=\prod_{i=1}^n(x-\Ga_i)$. Let $\Gb$ be the coefficient of $x$ in $f_V(x)$. Then the derivative $f'_V(x)$ is equal to $\Gb$. 

Let $D$ be the divisor  $\sum_{i=1}^nP_{i}$ and let $G=(k-1)\Pin$.  Consider the differential $\eta=\frac{\Gb}{f_V(x)}dx$. Then $(\eta)=-D+(n-2)\Pin$. Thus, $D-G+(\eta)=D-(k-1)\Pin-D+(n-2)\Pin=(n-k-1)\Pin\ge (k-1)\Pin=G$. So, we have $D+(\eta)\ge 2G$.

For each $1\le i\le n$, let $f_i(x):=\prod_{j\neq i}(x-\Ga_j)$. Then $(x-\Ga_i)f_i(x)=f_V(x)$. Taking the derivatives for the both sides gives $f_i(x)+(x-\Ga_i)f_i'(x)=f'_V(x)=\Gb$. This gives $f_i(\Ga_i)=\Gb$.
As $\eta=\frac1{x-\Ga_i}\times \frac{\Gb}{\prod_{j\neq i}(x-\Ga_j)}d(x-\Ga_i)$, we have $\res_{P_i}=\frac{\Gb}{\prod_{j\neq i}(\Ga_i-\Ga_j)}=\frac{\Gb}{f_i(\Ga_i)}=1$. Hence, $\res_{P_i}$ is a nonzero square in $\F_q$ for all $1\le i\le n$.  Hence, ${\bf res}(\eta)$ is equal to $\bu^2$ for some $\bu\in(\F_q^*)^n$. Finally, by Theorem \ref{thm:3.3} the code $\mC_L(D,G,\bu)$ is a $q$-ary $[n,k]$ self-orthogonal MDS code.
\end{proof}

Furthermore, we consider the case where $n$ divides $q-1$.

\begin{theorem}\label{thm:4.11}  Let $n|(q-1)$, then for
     any $1\le k\le {(n-1)/2}$,  there exists a $q$-ary $[n,k]$ self-orthogonal MDS code. 
\end{theorem}
\begin{proof} Let $V$ be a multiplicative subgroup of size $n$ in $\F_q^*$ and label the elements of $V$ by $\{\Ga_1,\dots,\Ga_n\}$. Then $x^n-1=\prod_{i=1}^n(x-\Ga_i)$. 

Let $D$ be the divisor  $\sum_{i=1}^nP_{i}$ and let $G=kP_0-\Pin$, where $P_0$ is the zero of $x$.   Consider the differential $\eta=\frac{nx^{n-1}}{x^n-1}dx$. Then $(\eta)=(n-1)P_0-D+(n-(n-1)-2)\Pin=(n-1)P_0-D-\Pin$. Thus, $D-G+(\eta)=D-kP_0+\Pin+(n-1)P_0-D-\Pin=(n-1-k)P_0 \ge (k+1)P_0\ge kP_0-\Pin=G$.

For each $1\le i\le n$, let $f_i(x):=\prod_{j\neq i}(x-\Ga_j)$. Then $(x-\Ga_i)f_i(x)=x^n-1$. Taking the derivatives for the both sides gives $f_i(x)+(x-\Ga_i)f_i'(x)=nx^{n-1}$. This gives $f_i(\Ga_i)=n\Ga_i^{n-1}$.
As $\eta=\frac1{x-\Ga_i}\times \frac{nx^{n-1}}{\prod_{j\neq i}(x-\Ga_j)}d(x-\Ga_i)$, we have $\res_{P_i}(\eta)=\frac{n\Ga_i^{n-1}}{\prod_{j\neq i}(\Ga_i-\Ga_j)}=1$. Hence, $\res_{P_i}$ is a nonzero square in $\F_q$ for all $1\le i\le n$, i.e., ${\bf res}(\eta)$ is equal to $\bu^2$ for some $\bu\in(\F_q^*)^n$. By Theorem \ref{thm:3.3} the code $\mC_L(D,G,\bu)$ is a $q$-ary $[n,k]$ self-orthogonal MDS code.
\end{proof}

Let us now  show nonexistence of $[n,n/2,n/2+1]$ self-dual MDS codes for the case where $n|(q-1)$, $n$ is even and $q\equiv 3\pmod{4}$. In fact, we have the following general result.

\begin{prop}\label{prop:4.13x}
If $q\equiv 3\pmod{4}$   and $k$ is odd, then there exist no $[2k,k,k+1]$ self-dual MDS codes over $\F_q$. In particular, if $n|(q-1)$ and $n$ is even, there exist no $[n,n/2,n/2+1]$ self-dual MDS codes over $\F_q$. 
\end{prop}
\begin{proof}
Suppose that there were a $[2k,k,k+1]$ self-dual MDS code over $\F_q$. Let $(I_k,P)$ be the systematic generator matrix of this code, where $I_k$ is the $k\times k$ identity matrix. Then we have $I_k+PP^T=O$. This gives $\det(P)^2=\det(PP^T)=\det(-I_k)=-1$. As $-1$ is a not a square when $q\equiv 3\pmod{4}$, we get a contradiction. 

As $q-1$ is equal to a product of $2$ with an odd number and $n|(q-1)$, $n$ is a product $2$ with an odd number $k$. Thus, a $[n,n/2,n/2+1]$ self-dual MDS code is a $[2k,k,k+1]$ self-dual MDS code. The desired result follows from the first part.
\end{proof}

Theorem \ref{thm:4.10} and Proposition \ref{prop:4.13x} completely solve the existence problem of self-orthogonal MDS code if $n|(q-1)$ except for 
the case where $q\equiv1\pmod{4}$, $n=q-1$ and $k=(q-1)/2$. 

\begin{theorem}~\label{thm:4.12} There exists a $q$-ary $[n,k]$ self-orthogonal MDS code if $k,n$ satisfies
\begin{itemize}
\item[{\rm (i)}]$(n+m)|q$ and $1\le k\le (n-m)/2$; or \item[{\rm (ii)}] $(n+m)|(q-1)$ and $1\le k\le (n-m-1)/2$.
\end{itemize}
\end{theorem}

\begin{proof} 
Assume that $(n+m)|q$. 
Let $q=p^\ell$ for a prime $p$ and an integer $\ell\ge 1$. Choose an $\F_q$ subspace $V$ of $\F_q$ with $|V|=n+m$ and label the elements of $V$ by $V=\{\Ga_1,\dots,\Ga_n,\Gb_1,\dots,\Gb_m\}$. Then we have the $\F_p$-linearized polynomial $f_V(x):=\prod_{i=1}^n(x-\Ga_i)\prod_{j=1}^n(x-\Gb_j)$. Let $\Gb$ is the coefficient of $x$ of $f_V(x)$. Then the derivative $f'_V(x)$ is equal to $\Gb$. 

Let $D$ be the divisor  $\sum_{i=1}^nP_{i}$ and let $G=(k+m-1)\Pin-\sum_{j=1}^mP_{\Gb_j}$.  Consider the differential $\eta=\frac{\Gb}{f_V(x)}dx$. Then $(\eta)=-D-\sum_{j=1}^mP_{\Gb_j}+(n+m-2)\Pin$. Thus, $D-G+(\eta)=D-(k+m-1)\Pin+\sum_{j=1}^mP_{\Gb_j}-D-\sum_{j=1}^mP_{\Gb_j}+(n+m-2)\Pin=(n-k-1)\Pin\ge (k+m-1)\Pin-\sum_{j=1}^mP_{\Gb_j}=G$ since $k\le(n-m)/2$.

For each $1\le i\le n$, let $f_i(x):=\frac1{x-\Ga_i}f_V(x)$. Then $(x-\Ga_i)f_i(x)=f_V(x)$. Taking the derivatives for the both sides gives $f_i(x)+(x-\Ga_i)f_i'(x)=f'_V(x)=\Gb$. This gives $f_i(\Ga_i)=\Gb$.
As $\eta=\frac1{x-\Ga_i}\times \frac{\Gb}{\prod_{j\neq i}(x-\Ga_j)\prod_{j=1}^n(x-\Gb_j)}d(x-\Ga_i)$, we have $\res_{P_i}(\eta)=\frac{\Gb}{\prod_{j\neq i}(\Ga_i-\Ga_j)\prod_{j=1}^n(\Ga_i-\Gb_j)}=1$. Hence, $\res_{P_i}(\eta)$ is a nonzero square in $\F_q$ for all $1\le i\le n$.  This means that ${\bf res}(\eta)$ is equal to $\bu^2$ for some $\bu\in(\F_q^*)^n$. Finally, by Theorem \ref{thm:3.3} the code $\mC_L(D,G,\bu)$ is a $q$-ary $[n,k]$ self-orthogonal MDS code.

Now assume that $(n+m)|(q-1)$.
Let $V$ be a multiplicative subgroup of $\F_q^*$ and label the elements of $V$ by $\{\Ga_1,\dots,\Ga_n,\Gb_1,\dots,\Gb_m\}$. Then $x^{n+m}-1=\prod_{i=1}^n(x-\Ga_i) \prod_{j=1}^m(x-\Gb_j)$. 

Let $D$ be the divisor  $\sum_{i=1}^nP_{i}$ and let  $G=(k+m)P_0-\sum_{j=1}^mP_{\Gb_j}-\Pin$, where $P_0$ is the zero of $x$.   Consider the differential $\eta=\frac{(n+m)x^{n+m-1}}{x^{n+m}-1}dx$. Then $(\eta)=-D-\sum_{j=1}^mP_{\Gb_j}+(n+m-1)P_0-\Pin$. Thus, $D-G+(\eta)=D-(k+m)P_0+\sum_{j=1}^mP_{\Gb_j}+\Pin-D-\sum_{j=1}^mP_{\Gb_j}+(n+m-1)P_0-\Pin=(n-k-1)P_0\ge (k+m)P_0-\sum_{j=1}^mP_{\Gb_j}-\Pin=G$.

For each $1\le i\le n$, let $f_i(x):=\frac1{x-\Ga_i}(x^{n+m}-1)=\prod_{i\neq j}(x-\Ga_i) \prod_{j=1}^m(x-\Gb_j)$. Then $(x-\Ga_i)f_i(x)=x^{n+m}-1$. Taking the derivatives for the both sides gives $f_i(x)+(x-\Ga_i)f_i'(x)=(n+m)x^{m+n-1}$. This gives $f_i(\Ga_i)=(n+m)\Ga_i^{m+n-1}$.
As $\eta=\frac1{x-\Ga_i}\times \frac{(n+m)x^{n+m-1}}{f_i(x)}d(x-\Ga_i)$, we have $\res_{P_i}(\eta)=\frac{(m+n)\Ga_i^{m+n-1}}{f_i(\Ga_i)}=1$. Hence, $\res_{P_i}(\eta)$ is a nonzero square in $\F_q$ for all $1\le i\le q=n$.  This means that ${\bf res}(\eta)$ is equal to $\bu^2$ for some $\bu\in(\F_q^*)^n$. Finally, by Theorem \ref{thm:3.3} the code $\mC_L(D,G,\bu)$ is a $q$-ary $[n,k]$ self-orthogonal MDS code.
\end{proof}

\begin{lemma}
 Assume that $q\equiv 1\pmod{4}$.  Let $p$ be the characteristic of $\F_q$. Let $V$ be  an $\F_p$-subspace  of $\F_q$.  Then the coefficient of $x$ of $f_V(x)=\prod_{\Ga\in V}(x-\Ga)$ is a square of $\F_q$. 
\end{lemma}
\begin{proof} We prove this by induction on the $\F_p$-dimension of $V$. 

For $\dim_{\F_{q}}(V)=1$, we have that $V$ is spanned by a nonzero element $\Gg\in\F_q$. Thus, $f_V(x)=\prod_{\Gl\in\F_p}(x-\Gl \Gg)=x^p-\Gg^{p-1}x$. As $-1$ is a square under the condition that $q\equiv 1\pmod{4}$ and $p-1$ is even, $-\Gg^{p-1}$ is a square. 

Suppose that it is true for $\ell-1$ for $\ell\ge 2$. Let $\dim_{\F_{q}}(V)=\ell$. Write $V$ as a direct sum of $\F_p$-vector space $U$ of dimension $\ell-1$ and $\F_p$-vector space $W$ of dimension $1$. Assume that $W$ is spanned by $\Gb\in\F_q\setminus U$. Then we have  
\begin{eqnarray*}
f_V(x)&=&\prod_{\Gl\in\F_p}\prod_{\Ga\in V}(x-\Ga+\Gl \Gg)=\prod_{\Gl\in\F_p}f_U(x+\Gl\Gb)\\
&=&\prod_{\Gl\in\F_p}(f_U(x)+\Gl f_U(\Gb))=f_U(x)^p-f_U(\Gb)^{p-1}f_U(x).\end{eqnarray*}
This implies that the coefficient of $x$ in $f_V(x)$ is equal to the product of $-f_U(\Gb)^{p-1}$ with the coefficient of $x$ in $f_U(x)$. Hence, it is a square by induction. The proof is completed.
\end{proof}

\begin{theorem}~\label{thm:4.14}  Assume that $q\equiv 1\pmod{4}$. Let $(n-1)|q$ and $n<q+1$, then for
    any $1\le k\le n/2$,  there exists a $q$-ary $[n,k]$ self-orthogonal MDS code. 
\end{theorem}

\begin{proof} 
Let $q=p^\ell$ for a prime $p$ and an integer $\ell\ge 1$. Choose an $\F_p$ subspace $V$ of $\F_q$ with $|V|=n-1$ and label the elements of $V$ by $V=\{\Ga_1,\dots,\Ga_{n-1}\}$. Then we have the $\F_p$-linearized polynomial $f_V(x):=\prod_{i=1}^{n-1}(x-\Ga_i)$. Let $\Gb$ is the coefficient of $x$ in  $f_V(x)$. Then the derivative $f'_V(x)$ is equal to $\Gb$. Choose $\Gg\in\F_q\setminus V$ (this is possible since $n-1<q$).

Let $D$ be the divisor  $\Pin+\sum_{i=1}^{n-1}P_{i}$. Let $G=(k-1)P_\Gg$.  Consider the differential $\eta=\frac{(x-\Gg)^{n-2}}{f_V(x)}dx$. Then $(\eta)=-D+(n-2)P_\Gg$. Thus, $D-G+(\eta)=D-(k-1)P_\Gg-D+(n-2)P_\Gg=(n-k-1)P_\Gg\ge (k-1)P_\Gg=G$.

For each $1\le i\le n-1$, let $f_i(x):=\prod_{j\neq i}(x-\Ga_j)$. Then $(x-\Ga_i)f_i(x)=f_V(x)$. Taking the derivatives for the both sides gives $f_i(x)+(x-\Ga_i)f_i'(x)=f'_V(x)=\Gb$. This gives $f_i(\Ga_i)=\Gb$.
As $\eta=\frac1{x-\Ga_i}\times \frac{(x-\Gg)^{n-2}}{\prod_{j\neq i}(x-\Ga_j)}d(x-\Ga_i)$, we have {$\res_{P_i}(\eta)=\frac{(\Ga_i-\Gg)^{n-2}}{\prod_{j\neq i}(\Ga_i-\Ga_j)}=\frac{(\Ga_i-\Gg)^{n-2}}{\Gb}$.} Since $n$ is even and $\Gb$ is a square, thus $\res_{P_i}(\eta)$ is a nonzero square in $\F_q$ for all $1\le i\le n-1$. Furthermore, $\res_{\Pin}(\eta)=-1$.
Hence, ${\bf res}(\eta)$ is equal to $\bu^2$ for some $\bu\in(\F_q^*)^n$. Finally, by Theorem \ref{thm:3.3} the code $\mC_L(D,G,\bu)$ is a $q$-ary $[n,k]$ self-orthogonal MDS code.
\end{proof}

\section*{Acknowledgments}
We acknowledge that Proposition \ref{prop:4.5} Theorems \ref{thm:4.7} and \ref{thm:12x} were developed with the assistance of GPT-6.2 Sol.  The author takes full responsibility for the content, verification, and correctness of all results presented in this work. We are also grateful to Ojas Mor, Vadim Genkin and Hongfeng Wu for drawing our attention to Conjectures 1 and 2 and pointing out corrections in the first version.


\begin{thebibliography}{99}

\bibitem{B12} S. Ball, “On sets of vectors of a finite vector space in which every subset of basis size is a basis,” J. Eur. Math. Soc., vol. 14, no. 3, pp. 733–748, 2012.



\bibitem{FF19} W. Fang and F. Fu, "New constructions of MDS Euclidean self-dual codes from GRS codes and extended GRS codes," IEEE Trans. Inf. Theory, vol. 65, no. 9, pp. 5574-5579, 2019.



\bibitem{FLLL19} X. Fang, K. Lebed, H. Liu, and J. Luo, "New MDS Euclidean self-dual codes over finite fields of odd characteristic," arXiv:1811.02802v9 [cs.IT], Sep. 2019.
\bibitem{FLL21} Xiaolei Fang, Meiqing Liu and Jinquan Luo,
``New MDS Euclidean Self-Orthogonal Codes"
IEEE Trans. Inf. Theory, vol. 61, no. 1, pp. 130 -137, 2021.

\bibitem{GK02} S. Georgion and C. Koukouvinos, "MDS Euclidean self-dual codes over large prime fields," Finite Fields and Their Appl., vol. 8, no. 4, pp. 455-470, 2002.

\bibitem{GG08} M. Grassl and T. A. Gulliver, "On self-dual MDS codes," in Proc. of ISIT, pp. 1954-1957, 2008.

\bibitem{G12} K. Guenda, "New MDS Euclidean self-dual codes over finite fields," Des. Codes Cryptogr., vol. 62, no. 1, pp. 31-42, 2012.

\bibitem{GKL08} T. A. Gulliver, J. L. Kim, and Y. Lee, "New MDS or near-MDS Euclidean self-dual codes," IEEE Trans. Inf. Theory, vol. 54, no. 9, pp. 4354-4360, 2008.

\bibitem{HK05} M. Harada and H. Kharaghani, "Orthogonal designs, self-dual codes and the Leech lattice," J. Combin. Designs, vol. 13, no. 3, pp. 184-194, 2005.

\bibitem{HK06} M. Harada and H. Kharaghani, "Orthogonal designs and MDS Euclidean self-dual codes," Australas. J. Combin., vol. 35, pp. 57-67, 2006.

\bibitem{JX17} L. Jin and C. Xing, "New MDS self-dual codes from generalized Reed-Solomon codes," IEEE Trans. Inf. Theory, vol. 63, no. 3, pp. 1434-1438, 2017.

\bibitem{KZT12} X. Kai, S. Zhu S, and Y. Tang, "Some constacyclic self-dual codes over the integers modulo \({2}^{m}\) ," Finite Fields and Their Appl. vol. 18, no. 2, pp. 258-270, 2012.

\bibitem{KL04a} J. L. Kim and Y. Lee, "MDS Euclidean self-dual codes," in Proc. of ISIT, pp. 1872-1877, 2004.

\bibitem{KL04b} J. L. Kim and Y. Lee, "Euclidean and Hermitian self-dual MDS codes over large finite fields," J. Combin. Theory, Series A, vol. 105, no. 1, pp. 79-95, 2004.

\bibitem{LLL19} K. Lebed, H. Liu, and J. Luo, "Construction of MDS Euclidean self-dual codes over finite field," Finite Fields and Their Appl., vol. 59, pp. 199-207, 2019.

\bibitem{LX2004} S. Ling and C. Xing, "Coding Theory: A First Course", Cambridge University Press, 2004.


\bibitem{NX01} H. Niederreiter and C.P. Xing,  {\it Rational Points on Curves over Finite Fields: Theory and Applications},
 LMS {\bf 285}, Cambridge, 2001.


\bibitem{P65}	V. Pless, {\it The number of isotropic subspaces in a finite geometry,} Atti Accad. Naz. Lincei Rend. Cl. Sci. Fis. Mat. Natur. 39 (8) (1965) 418-421.


\bibitem{St09}  H. Stichtenoth, {\it Algebraic Function Fields and Codes}, Graduate Texts in Mathematics {\bf 254}, Springer Verlag, 2009.
\bibitem{TV91} M. A. Tsfasman and S.G. Vl\u{a}du\c{t}, {\it Algebraic-Geometric Codes}, Dordrecht, The Netherlands: Kluwer, 1991.

\bibitem{TW16} H. Tong and X. Wang, "New MDS Euclidean and Herimitian self-dual codes over finite fields," Advances in Pure Mathematics., vol. 7, no. 5, 2016.

\bibitem{Y19} H. Yan, "A note on the construction of MDS Euclidean self-dual codes," Cryptogr. Commun., vol. 11, no. 2, pp. 259-268, 2019.

\bibitem{YC15} Y. Yang and W. Cai, "On self-dual constacyclic codes over finite fields," Des. Codes Cryptogr., vol. 74, no. 2, pp. 355-364, 2015.

\bibitem{ZF19} A. Zhang and K. Feng, "An unified approach on constructing of MDS self-dual codes via Reed-Solomon codes," IEEE Transactions on Information Theory, Vol. 66, Issue: 6, June 2020.



\end{thebibliography}
\end{document}